\documentclass[final,1p,times,authoryear]{elsarticle}

\usepackage{amssymb}
\usepackage{graphicx}
\usepackage{algorithm}
\usepackage{algorithmic}
\usepackage{multirow}
\usepackage{amsmath}
\usepackage{xcolor}
\usepackage{color}
\usepackage{graphics}
\usepackage{epsfig}
\usepackage{longtable}
\usepackage{array}
\usepackage{epsfig,graphicx,subfigure}

\usepackage{amsthm}

\journal{Computers and Operations Research}

\begin{document}

\begin{frontmatter}



\title{An Efficient Quasi-physical Quasi-human Algorithm for Packing Equal Circles in a Circular Container}


\author[add1]{Kun He, Hui Ye\footnote{Corresponding author. Email:yehui20080414@gmail.com},
Zhengli Wang\footnote{Corresponding author. Email:498195342@qq.com} }
\author[add2]{Jingfa Liu}

\address[add1]{School of Computer Science, Huazhong University of Science and Technology, Wuhan 430074, China}
\address[add2]{School of Computer \& Software, Nanjing University of Information Science \& Technology, Nanjing 210044, China}

\begin{abstract}
\normalsize This paper addresses the equal circle packing problem, and proposes an efficient Quasi-physical Quasi-human (QPQH) algorithm. QPQH is based on a modified Broyden-Fletcher-Goldfarb-Shanno (BFGS) algorithm which we call the local BFGS and a new basin hopping strategy based on a Chinese idiom: alternate tension with relaxation. Starting from a random initial layout, we apply the local BFGS algorithm to reach a local minimum layout. The local BFGS algorithm fully utilizes the neighborhood information of each circle to considerably speed up the running time of the gradient descent process, and the efficiency is very apparent for large scale instances. When yielding a local minimum layout, the new basin-hopping strategy is to shrink the container size to different extent to generate several new layouts. Experimental results indicate that the new basin-hopping strategy is very efficient, especially for a type of layout with comparatively dense packing in the center and comparatively sparse packing around the boundary of the container. We test QPQH on the instances of $n = 1,2,\cdots,320$, and obtain 66 new layouts which have smaller container sizes than the current best-known results reported in literature.
\end{abstract}

\begin{keyword}
Circle packing \sep neighborhood structure \sep container shrinking strategy \sep quasi-physical \sep quasi-human


\end{keyword}

\end{frontmatter}


\section{Introduction}
\label{Intro}
The circle packing problem (CPP) is concerned with arranging $n$ circles in a container with no overlap, which is of great interest to the industry and academia. CPP is encountered in a variety of real world applications, including apparel, naval, automobile, aerospace and facility layout \citep{castillo2008solving}.
CPP has been proven to be NP-hard \citep{demaine2010circle}, and it is difficult to find an exact solution in polynomial time even for some specific instances.
Researchers resort to heuristic methods that fall into two categories: the construction method and the optimization method.

The construction method can be described as packing circles one by one using some specific rules. There is one type of rules that fixes the container radius and only concerns where to feasibly place the circles in the container. Algorithms include Max Hole Degree (MHD) algorithm \citep{huang2003two}, self look-ahead search strategy \citep {huang2005greedy,huang2006new}, Pruned-Enriched-Rosenbluth Method (PERM) \citep{lu2008perm}, and beam search algorithm \citep{akeb2009beam}. The other type of rules adjusts the container radius along with the construction procedure. Algorithms include the Best Local Position (BLP) series \citep{hifi2004approximate,hifi2006strip,hifi2007dynamic,akeb2010hybrid}, and hybrid beam search looking-ahead algorithm \citep{akeb2010hybrid}.

Different to the construction method, the optimization method doesn't directly obtain a good solution, but iteratively improve the solution based on an ordinary initial solution. The great variety of optimization method can be further classified into quasi-physical, quasi-human algorithm \citep{wang2002improved,Liu2016A}, Tabu search and simulated annealing hybrid approach \citep{zhang2005effective}, population basin hopping algorithm \citep{addis2008efficiently}, simulated annealing algorithm \citep{m2009Packing}, formulation space search algorithm \citep{lopez2013packing}, iterated local search algorithm \citep{fu2013iterated,ye2013iterated,Liu2009An,Liu2015Heuristic}, etc. In 2015, there are two new algorithms published that yield excellent results: iterated Tabu search and variable neighborhood descent algorithm \citep{zeng2015iterated} and evolutionary computation-based method \citep{flores2015evolutionary}. 

We address the classic CPP, the Equal Circle Packing Problem (ECPP). In this section, we first recall mathematical methods for ECPP, which are the basis of researching ECPP. Then we concentrate on heuristics for ECPP, which are more effective for large scale ECPP. We summarize our work finally.

\subsection {Mathematical methods for ECPP}
As a classical type of CPP, ECPP is still difficult in the field of mathematics.
In the early stage of the study, the value of $n$ is relatively small, and researchers used mathematical analysis to not only find the optimal layout but also provide proofs on the optimality.
\cite{kravitz1967packing}, the first scholar working on ECPP, gave the layout for $n = 2, 3,\cdots, 19$ with the container radius, but he provided no proof of optimality. \cite{graham1968sets} proved the optimality for $n = 2, 3, 4, 5, 6, 7$.
\cite{pirl1969mindestabstand} proved the optimality for $n = 2, 3, 4, 5, 6, 7, 8, 9, 10$, and provided the layout for $n = 11, 12, \cdots,19$ at the same time.
\cite{goldberg1971packing} improved the layout for $n = 14, 16, 17$ that Pirl provided, and Goldberg also provided the layout for $n = 20$ for the first time.
\cite{reis1975dense} improved the layout for $n = 17$ based on Pirl's research, and he first gave the layout for $n = 21, 22, 23, 24, 25$.
\cite{melissen1994densest} proved the layout configuration optimality for $n = 11$, and \cite{fodor1999densest,fodor2000densest,fodor2003densest} proved the optimality for $n = 12, 13, 19$.
To summed up, only the optimality for $n = 2, 3, 4, 5, 6, 7, 8, 9, 10, 11, 12, 13, 19$ has been proved so far.

\subsection {Heuristics for ECPP}
Heuristics demonstrate their high effectiveness on ECPP. In this subsection, we introduce landmark heuristics for ECPP, and two key issues for solving ECPP.
\subsubsection{Landmark heuristics}
When $n$ is relatively large, it is very difficult to find the optimal layout and prove the optimality. Heuristic algorithms for ECPP can be very efficient to find optimal or suboptimal layouts. Although heuristics may not guarantee the theoretical optimality, they can find some layout where the container radius is very close to the theoretical minimum.

\cite{graham1998dense} did some early work and proposed two heuristic methods.
The first method is to simulate the repulsion forces. It transforms ECPP to a problem of finding the minimum on $\sum_{1\leq i\leq j\leq n} \left(\frac{\lambda}{\left \|S_i-S_j \right \|^2}\right)^m$, where 
${S_1,S_2,\cdots,S_n}$ correspond to the coordinates for the set of circle centers in the container,
$\left \|S_i-S_j \right \| \geq 2$, $\lambda$ the zoom factor, and $m$ a large positive integer. For such objective function, we can use some existing methods such as gradient descent to find a layout with the local minimum value.
The second method is the billiards simulation. This is a quasi-physical method that regards the circle items as billiards. This algorithm starts with a small billiard radius and randomly assigns an  initial movement direction for each billiard. Then a series of collision motion occurs in the circular container. During the process movement, the authors slowly increase the sizes of the billiards. By repeatedly running the algorithm, it is possible to find the global optimal solution.
By the comprehensive use of repulsion forces and billiards simulation, Graham found the near-optimal layout for $n = 25, 26, \cdots, 65$.

There are many followup work based on heuristics. Here we highlight several landmark works.
\cite{akiyama2003maximin} used a greedy method to find local optimal solution. Their algorithm continuously improves the current layout by randomly moving one circle until the moving reaches an iteration limit (e.g. 300000). By repeatedly running the greedy method, Akiyama found much denser layouts for $n = 70, 73, 75, 77, 78, 79, 80$.
\cite{grosso2010solving} assumed that ECPP has the characteristic of ``funneling landscape", and used a monotone hopping strategy to look for the ``funnel bottom". In order to solve the funnel problem, they used the population hopping strategy to enhance the diversity of the layout. They found a number of denser layout schemes for $66\leq n\leq100$.
\cite{huang2011global} proposed a global optimization algorithm using a quasi physical model. They proposed two new quasi physical strategies and
found 63 denser layouts among the 200 instances for $n = 1, 2, \cdots, 199, 200$.

\subsubsection{Two key issues}
There are two key issues to solve the ECPP. First, how can we optimize a random or given layout so that it is more likely to reach the local optimum layout. Second, when we reach a local optimal layout which is not feasible, that is, there exists overlapping among some circles, we need to use some strategy to jump out of the local minimum layout and reach a new layout that inherits the advantages of the previous local optimum. Then we could continue the local optimization to reach another local minimum, and we aim to eventually obtain an optimal or near-optimal layout.

For the local optimization method, the repulsion forces and billiards simulation of \cite{graham1998dense}, the monotone hopping strategy of \cite{grosso2010solving}, and the elastic force movement of \cite{huang2011global} described above all fall into this category.
There are also other good methods, like TAMSASS-PECS method \citep{szabo2005global},  nonlinear optimization method \citep{birgin2005optimizing} and reformulation descent algorithm \citep{mladenovic2005reformulation}. Each of these algorithms has its own advantages, depending on the number of circles and the container shapes (square, circle, rectangle, or polygon).

There are diverse methods for the Basin-hopping strategy. For example, the small random perturbation method \citep{addis2008disk} formed a new layout by moving several circles in the local optimal layout to some random places. However, due to purely randomness, this method may destroy the holistic heredity.
\cite{huang2011global} considered three kinds of forces, elastic force, attractive force and repulsive force, to promote the entire layout to a new form. They used three parameters \emph{$c_1$}, \emph{$c_2$} and \emph{steps} to control the strength of the attractive force, the strength of the repulsive force and the duration time of the abrupt movement.
\cite{zeng2015iterated} proposed another strategy for moving random circles to vacant places in the container. By dividing the entire container into square grids, Zeng et al. regarded the vacant point with large vacant degree as candidate insertion place for the center of the ``jumping circle", which can improve the current layout to a certain extent.

\subsection {Our work}
We propose an efficient Quasi-physical Quasi-human (QPQH) algorithm for solving ECPP. We adopted the physical model \citep{huang2011global} popularly used for solving CPP. And through the establishment of the physical model, we look for a minimum of the objective function using the classical Quasi-Newton method, the Broyden-Fletcher-Goldfarb-Shanno (BFGS) algorithm \citep{Liu1989On}. To speed up the computation without losing much accuracy, we fully utilize the neighborhood structure of the circles, and propose a local BFGS algorithm. We also propose a new Basin-hopping strategy by shrinking the radius of the container. In the proposed QPQH, we iteratively apply the local BFGS algorithm to achieve a new layout after a certain number of continuous optimization iterations, and apply the Basin-hopping strategy to jump out of the local minimum. Experiments on 320 ($n=1,2,\cdots,320$) ECPP instances demonstrate the effectiveness of the proposed method.

\section{Problem Formulation}
\label{Pro}
\begin{figure}
\centering
\includegraphics[width=0.5\textwidth]{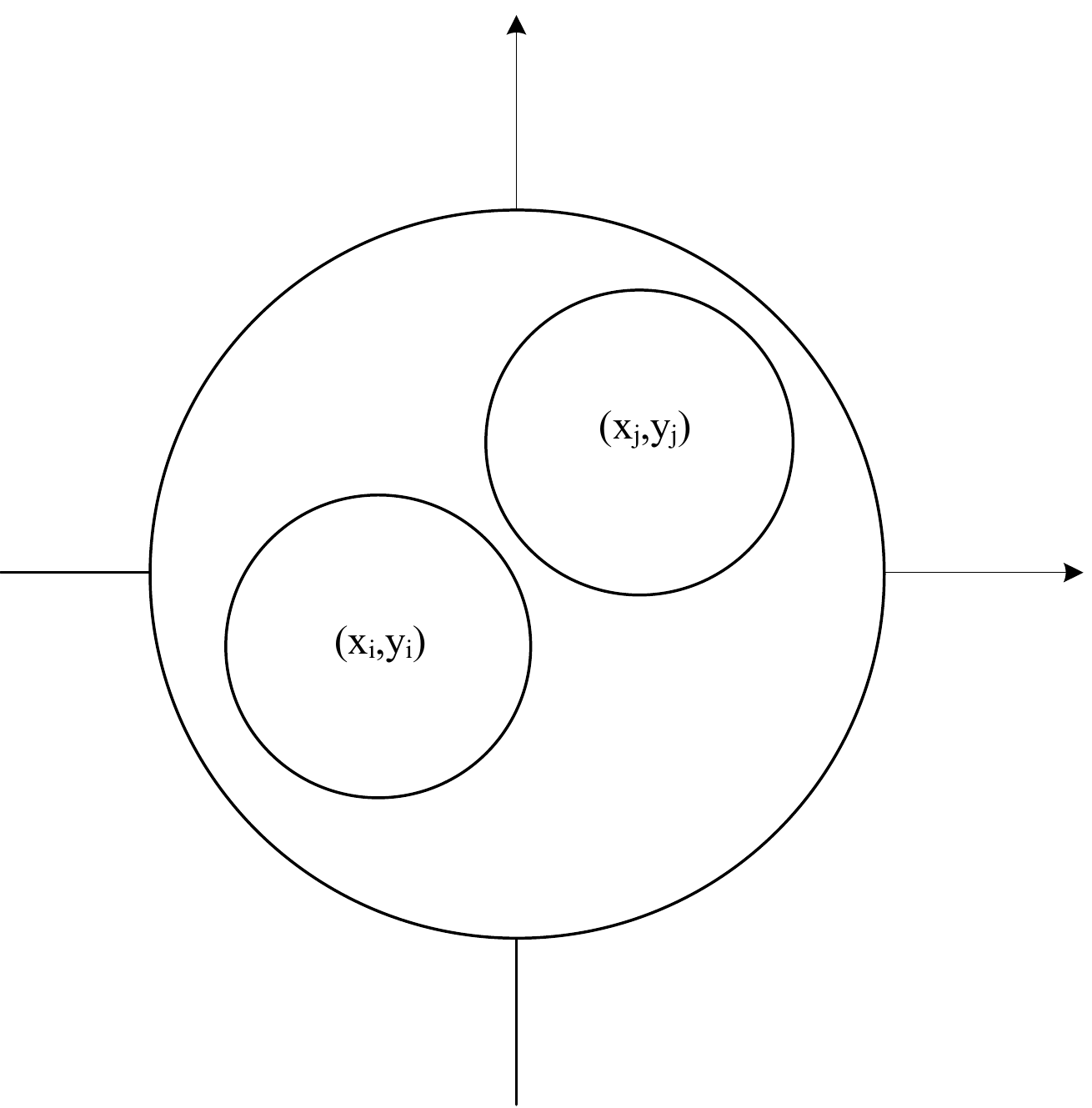}
\caption{The coordinate system.}
\label{fig.1}
\end{figure}

The Equal Circle Packing Problem (ECPP) can be described as
packing $n$ unit circle items into a circular container.
We need to ensure no overlap between any two circles and the
radius of the container, $R$, is minimized. By setting a Cartesian coordinate system with
its origin (0,0) located at the container center, as shown in Figure \ref{fig.1}, and denoting the coordinate for the center of circle $i$ $(i = 1,2,\cdots ,n)$ as $(x_i,y_i)$, we want to find a layout configuration $X=(x_1,y_1,\cdots,x_n,y_n)$ such that the container's radius is minimized.
\begin{align}
minimize &\qquad R  \notag \\
s.t. &\qquad (1)\qquad \sqrt{x_i^2+y_i^2}\leq R-1   \notag  \\
     &\qquad (2)\qquad \sqrt{{(x_i-x_j)}^2+{(y_i-y_j)}^2}\geq 2   \notag
\end{align}
where $i, j = 1,2,\cdots,n$ and $i \neq j.$

\section{Algorithm Description}
\label{Algdes}
Finding the minimum radius $R$ of the container is an optimization problem. As $R$ is unknown, it is difficult to directly find the minimum $R$. We can transform the problem into a decision problem, that is, for a fixed $R$, whether we can find a layout to meet the above conditions (1) and (2). If the response is ``yes", then we will continue reduce the value of $R$, until we can not find any feasible layout on a smaller $R$.

This section describes in detail the algorithm we proposed. We first introduce a popular physical model for solving this problem, and give a brief introduction to the local search algorithm, the classic BFGS algorithm. Then, we propose the local BFGS algorithm, for each circle we form an adjacency list to record the current overlapping circles as well as possible overlapping circles in the near future. Next, we propose a new basin-hopping strategy, and the global search algorithm combining the local BFGS algorithm and the basin-hopping strategy. And we introduce a post-processing procedure, the container adjustment procedure, to yield a container radius as small as possible while maintaining feasibility. Finally, we add an overall pseudo-code integrating all sub-algorithms in QPQH.

\subsection {Physical model}
Similar to previous quasi-physical methods \citep{ye2013iterated,Huang1982A,He2013A,He2015An}, we regard the circle items as elastic items and imagine the container as a rigid circular container. If an elastic object embeds itself in another object, it will be deformed and the deformation will cause some elastic potential energy. Based on the Cartesian coordinate system, we have the following definitions.\\
\textbf{Definition 1} (\emph{Overlapping Depth}). There are two kinds of overlaps, item-item overlap and item-container overlap.
The \emph{overlapping depth} between item $i$ and the container is $d_{oi}$.

 \begin{equation}
  d_{oi} = max (\sqrt{x_i^2+y_i^2}+1-R, 0)
 \end{equation}

And the \emph{overlapping depth} between item $i$ and item $j$ ($i \neq j$) is $d_{ij}$.
 \begin{equation}
  d_{ij} = max (2-\sqrt{{(x_i-x_j)}^2+{(y_i-y_j)}^2}, 0)
 \end{equation}
\textbf{Definition 2} (\emph{Elastic Energy}). According to the elastic mechanics, the elastic potential energy between two smoothy elastic objects is proportional to the square of the embedded depth. The elastic energy between item $i$ and $j$ $(j \neq i)$ is defined as following, where item 0 indicates the hollow object of the container.
\begin{align}
& U_{ij} = d_{ij}^2  \label{eq:rel3}
\end{align}
And the elastic energy $U_i$ of item $i$ is defined as:
\begin{align}
& U_i = \sum_{j=0,j \neq i}^n U_{ij}  \label{eq:rel4}
\end{align}
The total elastic energy $U(X)$ for the whole system $X = \{x_1,y_1,...,x_n,y_n\}$ is defined as:
\begin{align}
& U = \sum_{i=1}^n U_{i}   \label{eq:rel5}
\end{align}

\subsection {The BFGS continuous optimization algorithm}
With the physical model established, the potential energy function $U(X)$ is defined and it has the following properties:

(1) For all $X \in (-\infty, +\infty)^{n}$, $U(X) \geq 0$;

(2) $X$ is a feasible packing layout if and only if $U(X) = 0$.

Thus, the packing problem is transformed into the optimization problem of finding the minimum value on function $U(X)$. We use the classical Quasi-Newton method, the BFGS algorithm, for the continuous optimization of finding local minimums.

The basic idea of the BFGS algorithm is as follows. From a layout configuration $X_k$, we construct an approximate Hessian matrix $H_k$ using information of the objective function $U$ and the gradient $g_k$. Thus we can get a search direction $d_k$ and a search length $\lambda_k$, and generate a new layout configuration $X_{k+1}$. We continue the process iteratively until reaching a minimum layout configuration.

The gradient $g_k$ is defined by Eq. (\ref{eq:rel6}), and the search step length $\lambda_k$ is defined by Eq. (\ref{eq:rel7}). $\lambda_k$ is a positive real number in order to reach the minimum value of the elastic energy $U$ at step $k$. $X_k$ corresponds to a $2n-coordinate$ vector, and
$d_k$, defined by Eq. (\ref{eq:relb}), is the search direction at step $k$ 
\begin{align}
& g_k = \left(\frac{dU_k}{dx_1},\frac{dU_k}{dy_1},\cdots,\frac{dU_k}{dx_n},\frac{dU_k}{dy_n}\right)  \label{eq:rel6} \\
& \lambda_k = \arg{\min_{\lambda \in R^+}U(x_k + \lambda d_k)}  \label{eq:rel7}
\end{align}

Hessian matrix is a symmetric matrix composed of the second derivatives of $U(X_k)$, and $H$ is an approximate Hessian matrix of size $2n$ by $2n$. It is updated by Eq. (\ref{eq:rela}), where $I$ is the unit matrix, $s_k$ is the change quantity of $X_k$, and $y_k$ is the change quantity of $g_k$. The pseudo code of the BFGS algorithm is described in Algorithm \ref{alg:BFGS}.
\begin{align}
& H_{k+1} = \left(I  - \frac{s_k y_k^T}{y_k^T s_k}\right)H_k\left(I - \frac{y_k s_k^T}{y_k^T s_k}\right) + \frac{s_k s_k^T}{y_k^T s_k} \label{eq:rela}
\end{align}

\begin{align}
& d_k = -H_k * g_k \label{eq:relb}
\end{align}

\begin{algorithm}[!hbt]
\small
\caption{The BFGS algorithm($X$, $R$)}
\label{alg:BFGS}
\begin{algorithmic}[1]
\REQUIRE ~~\\
    An initial layout ($X$, $R$)
\ENSURE ~~\\
    A locally optimal packing layout ($X^*$)
    \STATE $H_0 \leftarrow I$; //$H_0$ is the initial approximate Hessian matrix, and $I$ is the unit matrix
    \STATE $k \leftarrow 0 $; //$k$ is the iteration step
\WHILE {$k \leq MaxIterations$}
    \STATE calculate the gradient $g_k$ by Eq. (\ref{eq:rel6});
    \STATE calculate the search direction $d_k \leftarrow -H_k*g_k$;
    \STATE calculate the search step length $\lambda_k$ by Eq. (\ref{eq:rel7});
    \STATE $s_k \leftarrow \lambda_k*d_k$, $X_{k+1} \leftarrow X_k+s_k$;
    \IF {$U < 10^{-20} $ or $\left \| g_k \right \| < 10^{-10}$}
       \RETURN the current layout ($X^*$);
    \ENDIF
    \STATE $y_k \leftarrow g_{k+1}-g_k$;
    \STATE $H_{k+1} \leftarrow \left(I  - \frac{s_k y_k^T}{y_k^T s_k}\right)H_k\left(I - \frac{y_k s_k^T}{y_k^T s_k}\right) + \frac{s_k s_k^T}{y_k^T s_k}$;
    \STATE $k \leftarrow k+1$;
\ENDWHILE
\end{algorithmic}
\end{algorithm}

\subsection {The local BFGS algorithm}
In the BFGS algorithm, when we use Eq. (\ref{eq:rel5}) and Eq. (\ref{eq:rel6}) to calculate $U$ and $g_k$, for each circle $i$, we need to calculate its distance to each of the other $n-1$ circles in order to judge whether there exist some overlaps. In the actual procedure, however, a circle usually intersects with approximately six other circles when there exist only slight overlaps. Ideally, if three equal circles are tangent to each other, the three lines connecting the pairwise circle centers construct an equilateral triangle. The angle between two lines is 60 degrees. As the maximum degree of the central angle in a circle is 360 degree. So there are at most six other circles tangent to a circle if there exists no overlap.

\begin{figure}[htb]
\centering
\includegraphics[width=0.5\textwidth]{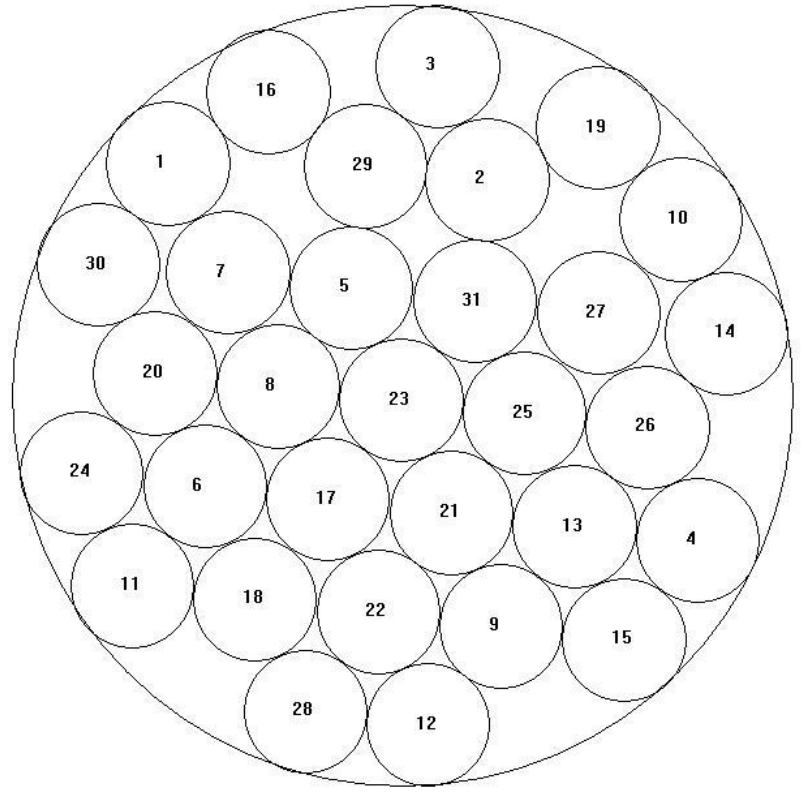}
\caption{The local minimum layout when $n$ = 31.}
\label{fig.2}
\end{figure}

\begin{figure}[htb]
\centering
\includegraphics[width=0.4\textwidth]{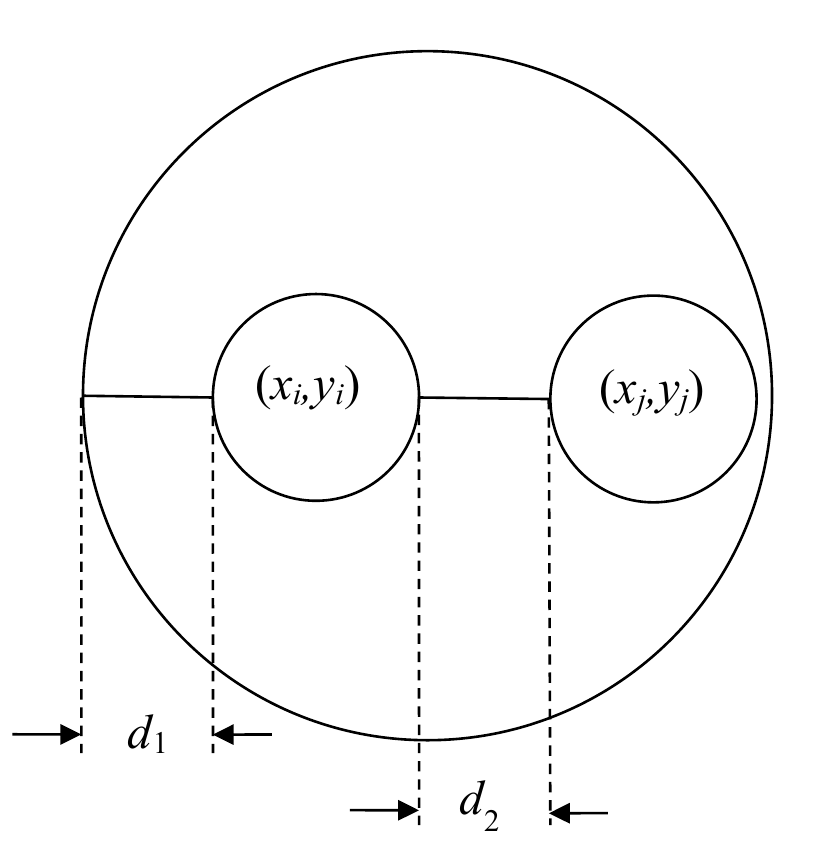}
\caption{Distance $d_1$ between circle $i$ and the container, and distance $d_2$ between circles $i$ and $j$.}
\label{fig.3}
\end{figure}

As illustrated in Figure \ref{fig.2}, there will be tiny changes for the adjacency relationship of the circles after certain iterations of the BFGS algorithm.
So during the calculation procedure, for each circle in the system, we record its overlapping circles at present and possible overlapping circles in the near future in an adjacency list.
Specifically, for a circle $i$ located at $(x_i, y_i)$, the container is regarded as an adjacency object if Eq. (\ref{eq:rel8}) holds;
for a circle $j$ located at $(x_j,y_j)$ is regarded as an adjacency object if Eq. (\ref{eq:rel9}) holds.
We set $d_1 = 1, d_2 = 1$, which means an object (circle or container) is considered as adjacent if the distance from its boundary to the boundary of the current circle is no greater than 1, as illustrated in Figure \ref{fig.3}.
\begin{align}
& \sqrt{x_i^2+y_i^2} + 1 - R \leq d_1  \label{eq:rel8} \\
& \sqrt{{(x_i-x_j)}^2+{(y_i-y_j)}^2} - 2 \leq  d_2  \label{eq:rel9}
\end{align}

We only check the overlapping status for circles in the adjacency list in the local BFGS algorithm.
And we will update the adjacency list for each circle after running the local BFGS for $l$ iterations. If $l$ is too large, then there could be some big error for the overlapping approximation. If $l$ is too small, then we need to frequently recalculate the distance for all other circles for each circle and could not really speed up the BFGS optimization. We set $l$ = 10 as a tradeoff selection.

\begin{figure}[htb]
\centering
\includegraphics[width=0.5\textwidth]{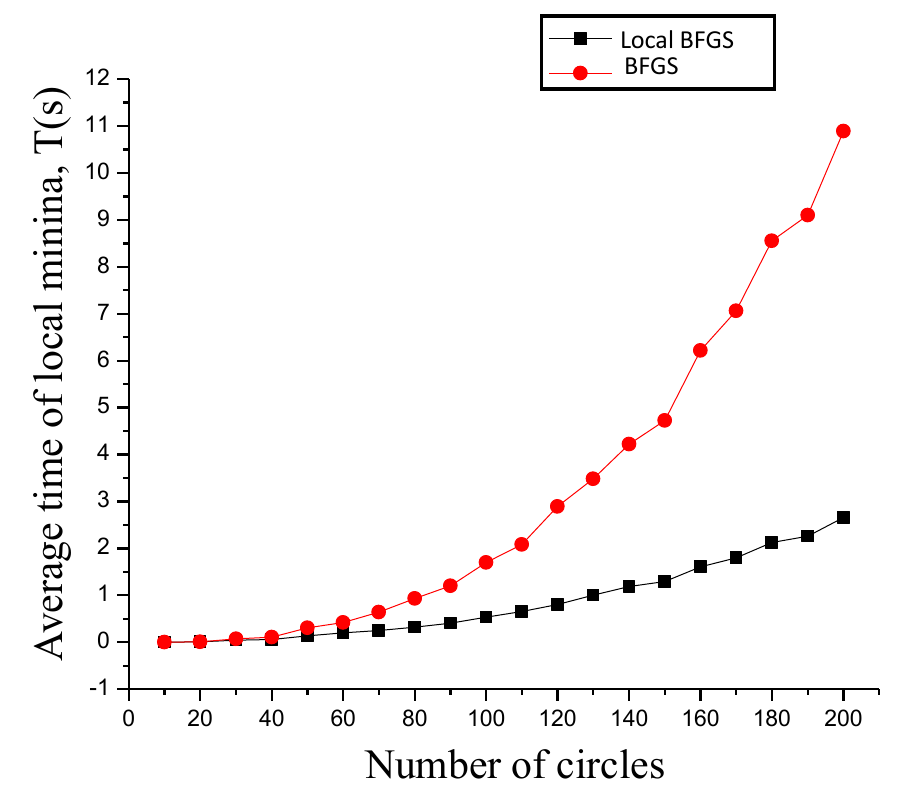}
\caption{Improvement on the average minimum time of the local BFGS method.}
\label{fig.12}
\end{figure}

We experimentally test the effectiveness of the local BFGS algorithm on 20 instances, $n=10,20,\cdots, 200$. We set the radius of the container as the current known minimum radius, and record the time starting from a random initial layout to a local minimum layout. We calculate each instance for 1000 times, record the average running time $T$ of reaching the local minimum, and compare with the original BFGS algorithm.
The comparison result is shown in Figure {\ref{fig.12}}. With the increasement on the scale $n$, the average time $T$ of the two methods both increases. However, the running time of the BFGS algorithm increases much faster. In the case of $n=200$, the value $T$ of the BFGS algorithm is about 4 times as compared with the value $T$ of the local BFGS algorithm.

When using Eq. (\ref{eq:rel5}) to compute $U$ and Eq. (\ref{eq:rel6}) to compute $g_k$, we only consider the adjacency circles, the time complexity reduces from $O(N^2)$ to $O(N)$. But the total complexity is still $O(N^2)$, as we need to update the approximate Hessian matrix which is in $n$ by $n$ scale.
Nevertherless, the computational efficiency shows great improvement for the local optimization, especially for relatively large instances (such as $n \geq 200$), as shown in Figure \ref{fig.12}.

\subsection {Basin-hopping procedure}
When reaching a local minimum by the local BFGS, we shrink the container radius as the Basin-hopping strategy.
We fix the original coordinate of each circle and reduce the container radius by a factor of $\beta$ ($0 < \beta < 1$): $R_0 = \beta R_0$.
$\beta$ is defined as $\beta = 0.3 + 0.035m$ and $m$ varies from 0, 1, 2, to 19 such that $\beta$ varies from 0.30 to 0.965 to generate diverse layouts. 0.3 is just an empirical setting and $0.035 = \frac{0.7}{20}$. Other alternative setting also works. For instance, if you choose the shrinking ratio to start from 0.4 and want to generate 20 new layouts, then you should set $\beta = 0.4 + 0.03m$ as $0.03 = \frac{0.6	}{20}$, and $R$ shrinks from $0.4R$ to $0.97R$.

\begin{figure}[htb]
\centering
\includegraphics[width=0.7\textwidth]{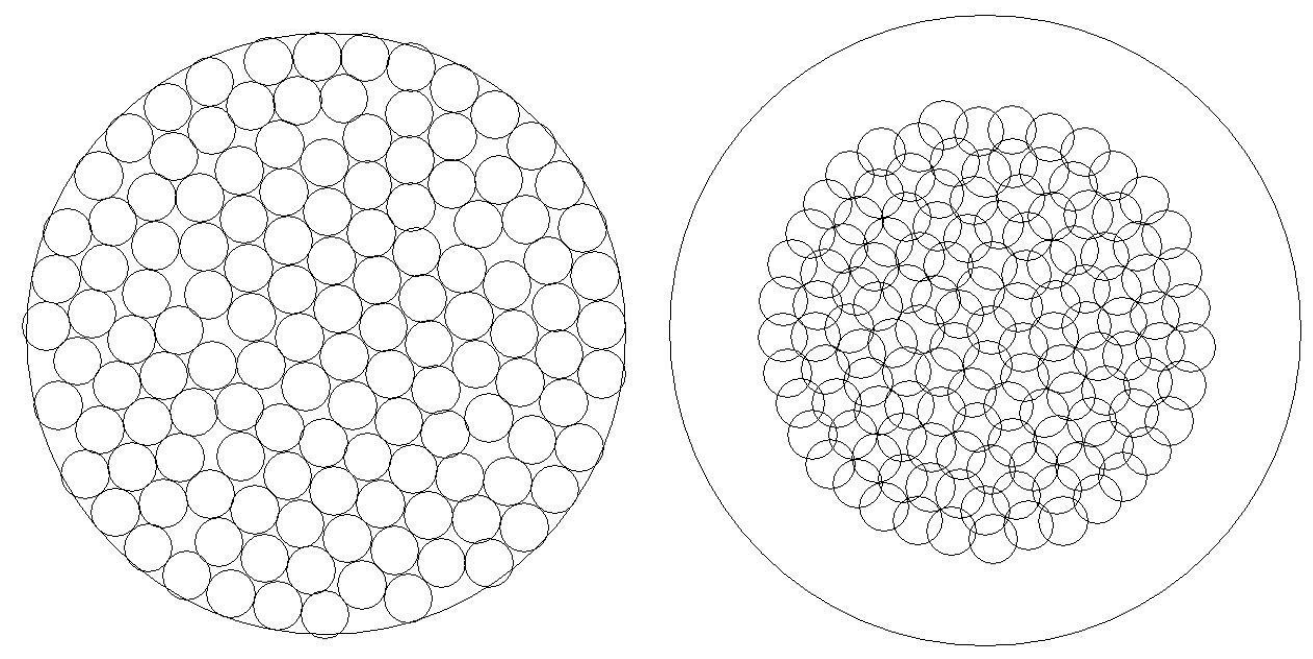}
\caption{The layout before and after the Basin-hopping procedure for $n$ = 130.}
\label{fig.4}
\end{figure}

We then run the local BFGS algorithm for $h$ iterations ($h$ is a random number ranging from 50 to 100) to reach a new layout. Figure \ref{fig.4} represents the layout before and after the Basin-hopping procedure.
The pseudo-code of the Basin-hopping procedure is shown in Algorithm \ref{alg:Bh}.
We reduce the container radius $R_0$ to $\beta R_0$, and run the local BFGS for $h$ iterations regarding the current minimum layout as an initial layout. In this way, we have 20 new layouts as the basin hopping results.

\begin{algorithm}[!hbt]
\small
\caption{Basin-hopping procedure$(X,R)$}
\label{alg:Bh}
\begin{algorithmic}[1]
\REQUIRE ~~\\
    A local minimum layout $(X,R)$
\ENSURE ~~\\
    20 new layouts.
    \STATE $R_0 \leftarrow R$;
\FOR {$m \leftarrow 0$ to $19$}
    \STATE $\beta \leftarrow 0.3 + 0.035m$;
    \STATE $R \leftarrow \beta R_0$;
    \STATE $h \leftarrow $a random integer in [50, 100];
    \STATE run the local $BFGS(X,R)$ for $h$ iterations to reach a new layout $X_m$; 
\ENDFOR
    \RETURN the 20 new layouts $(X_m)$($m = 0,1,\cdots,19$).
\end{algorithmic}
\end{algorithm}

We experimentally test the effectiveness of the Basin-hopping procedure. We test the instances for $n=50,51,\cdots,70$ to compare the Basin-hopping procedure with random initialization method. The radius of the container is set as the current known minimum radius. We compare the number of iterative steps needed to find a feasible layout using the local BFGS algorithm.

\begin{figure}[htb]
\centering
\includegraphics[width=0.55\textwidth]{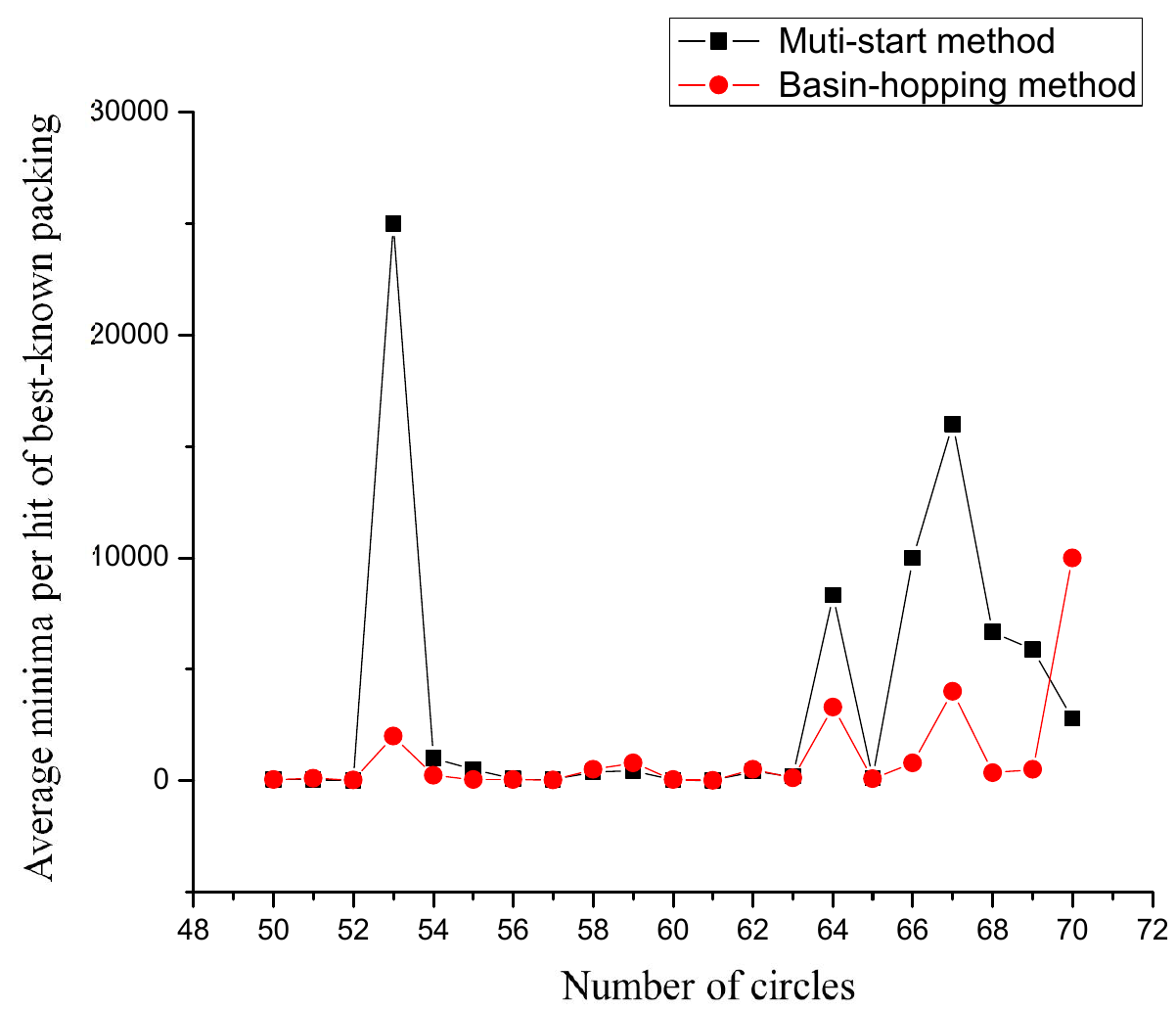}
\caption{Comparison of the shrinking Basin-hopping procedure with random the initialization method.}
\label{fig.13}
\end{figure}

The calculation results are shown in Figure \ref{fig.13}. It shows that, the Basin-hopping procedure is apparently better in the cases of $n$ = 53, 54, 55, 66, 67, 68, 69; In the case of $n=70$, the random initial method is better. In other cases, the effectiveness of two methods are almost the same. On the average, the shrinking strategy exhibits better performance.

\begin{figure}[H]
\centering
\includegraphics[width=0.6\textwidth]{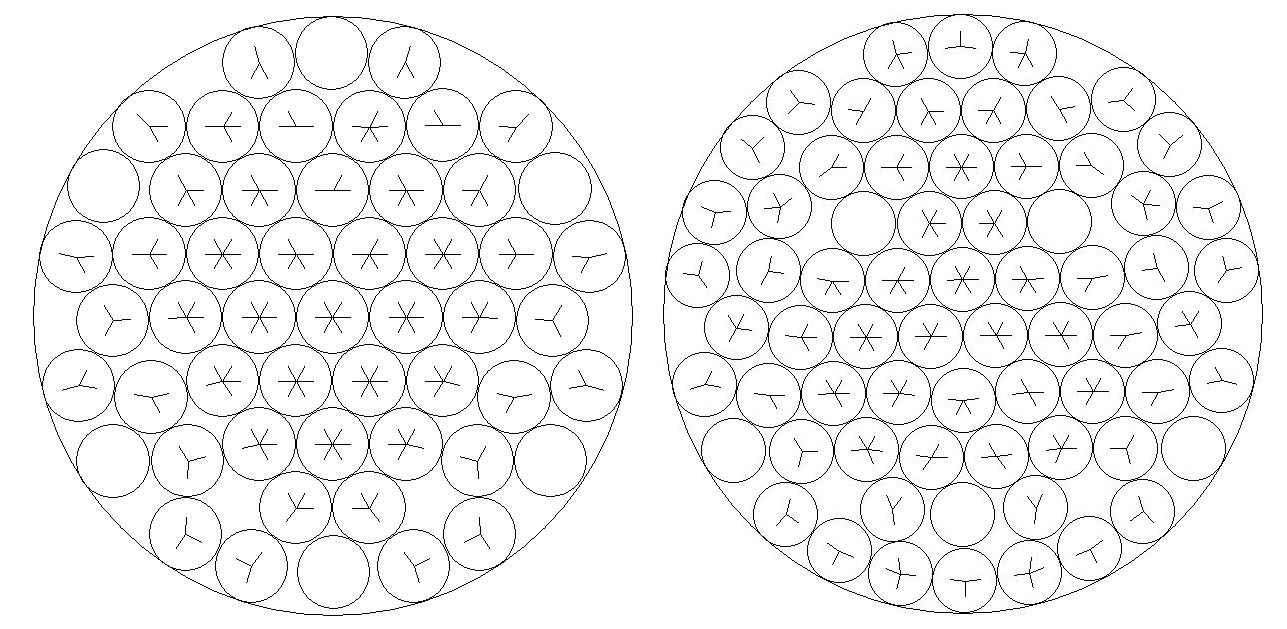}
\caption{The resulting minimum layouts for $n$ = 53 and $n$ = 68.}
\label{fig.14}
\end{figure}

In the shrinking Basin-hopping procedure, the container's radius decreases dramatically. After several iterations using the local BFGS algorithm, all circles contract to the container center. Then we recover the container radius and there exists a large vacant space in the outer layer of the container. In the iterative process of the local BFGS, all circles move outwards as a whole. While the outer space is large and the inner space is small, the result of the final movement is that inner circles are dense and outer circles are sparse. Therefore, the Basin-hopping procedure has better influence for instances with dense inner packing and sparse outer packing, which are the most cases for the near-optimal or dense packing. We show two typical examples with the above characteristics in Figure \ref{fig.14}.

Combining the local BFGS algorithm with the Basin-hopping procedure, we present the global search algorithm in Algorithm \ref{alg:global}. We first run the local BFGS with a random initial layout $X$. When reaching a local minimum layout, we run the Basin-hopping procedure with the local minimum layout to generate 20 new layouts. Then we run the local BFGS on the 20 new layouts respectively. The algorithm will terminate once it finds a feasible layout. If none of the 20 new layouts becomes feasible, we will randomly generate an initial layout again and repeat the calculation until a time limit $t_0$ is reached.

\begin{algorithm}[!hbt]
\small
\caption{The global search algorithm}
\label{alg:global}
\begin{algorithmic}[1]
\REQUIRE ~~\\
   A container radius $R$
\ENSURE ~~\\
    A feasible or stuck layout $(X,R)$
\WHILE {$U > 10^{-20}$ and the timeout limit $t_0$ is not reached }
    \STATE randomly generate an initial layout $X$;
    \STATE run the local BFGS algorithm($X$, $R$);
    \IF {$U < 10^{-20}$}
        \STATE break;
    \ENDIF
    \STATE run Basin-hopping $(X, R)$ to generate 20 new layouts and store them in array $C$;
    \FOR {$i = 1$ to 20}
       \STATE run the local BFGS algorithm$(C[i],R)$;
       \IF {$U < 10^{-20}$}
         \STATE break;
       \ENDIF
    \ENDFOR
\ENDWHILE
\RETURN the current layout($X$, $R$).
\end{algorithmic}
\end{algorithm}

\subsection {The container adjustment procedure}
In the global search algorithm, for the new layout after the Basin-hopping, the radius and location of the container are constant, and circles are centralized around the center of the container. We then continue the optimization with the local BFGS algorithm and hopefully to reach a feasible layout. During this procedure, viewing from the layout picture, most circles are moving from the container's center to the border. If the termination condition
$U < 10^{-20}$ is satisfied while the circles haven't been tangent to the border of the container, it is obviously that the radius of the container can be reduced until the circles are tangent to the border of the container. So as a post processing, we use the container adjustment strategy to decrease the radius of the container.

If the global search algorithm returns a feasible layout, we hope to maintain the relative locations of the circles and further decrease the radius of the container to get a more compact layout. Here we use the dichotomy method to find the minimal radius $R_{min}$ of the container from the current layout.
We update $R_{bestknown}$ to $R_{min}$, and use the global search algorithm for a new round of search.

\begin{algorithm}[!htbp]
\small
\caption{Container adjustment algorithm}
\label{alg:adjust}
\begin{algorithmic}[1]
\REQUIRE ~~\\
    Current best minimum layout $(X,R)$
\ENSURE ~~\\
    New feasible layout $(X,R_{min})$
    \STATE $R_0 \leftarrow R$;
    \STATE $i \leftarrow -1 $;
\REPEAT
    \STATE $R_{upperbound} \leftarrow R_0$;
    \STATE $i \leftarrow i+1$;
    \STATE $R_0 \leftarrow R_{upperbound} - 10^{-10} \times 10 ^ i$;
    \STATE run the local BFGS algorithm($X$, $R_0$);
\UNTIL {$(X_0,R)$ is not feasible};
    \STATE $R_{lowerbound} \leftarrow R_0$;
\REPEAT
    \STATE $R_{min} \leftarrow (R_{upperbound} + R_{lowerbound})/2 $;
    \STATE run the local BFGS algorithm($X$,$R_{min}$);
    \IF {$(X,R_{min})$ is feasible}
        \STATE $R_{upperbound} \leftarrow R_{min}$;
    \ELSE \STATE $R_{lowerbound} \leftarrow R_{min}$;
    \ENDIF
\UNTIL {$|R_{upperbound} - R_{lowerbound}| <= 10^{-10}$};
\RETURN the current feasible layout $(X,R_{min})$.
\end{algorithmic}
\end{algorithm}

\begin{figure}[htb]
\centering
\includegraphics[width=0.8\textwidth]{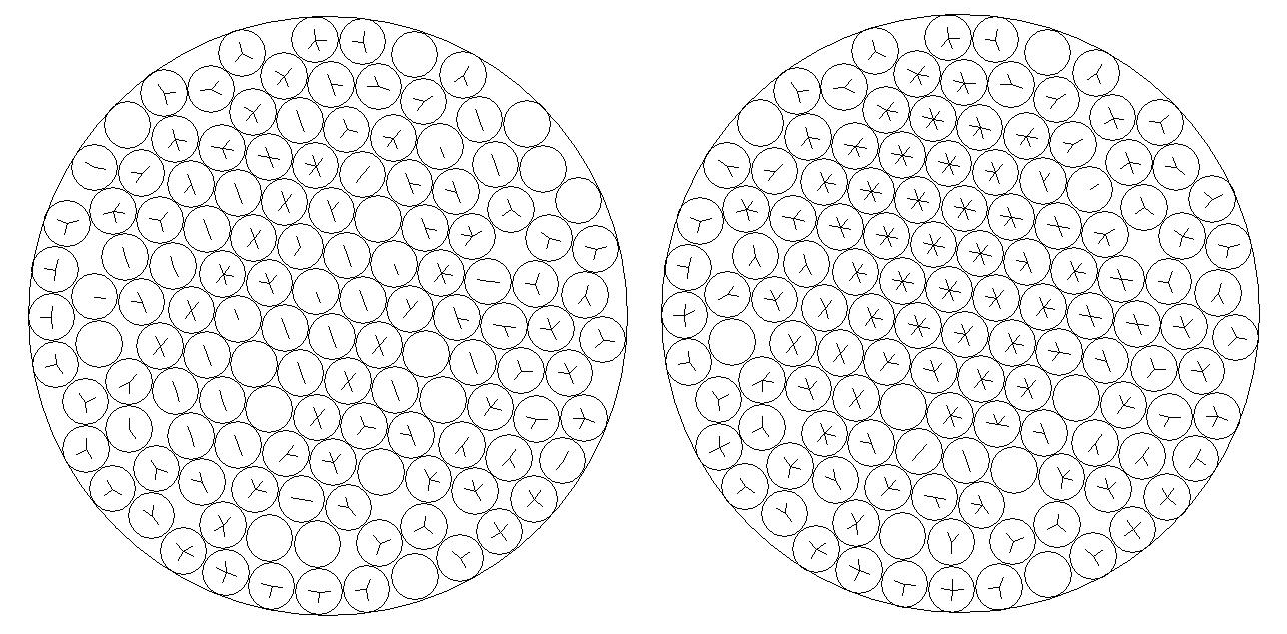}
\caption{The layout before and after the container adjustment procedure for $n = 130$.}
\label{fig.5}
\end{figure}

For the instance of $n=130$, $R_{bestknown} = 12.6023189367$. The global search algorithm finds the global minimum layout, as shown at the left of Figure {\ref{fig.5}}. After the container adjustment procedure, the container radius is changed to $R_{min} = 12.6017746136$, which is reduced by about $6\times10^{-4}$, as shown at the right of Figure {\ref{fig.5}}.

\subsection {The Efficient Quasi-physical Quasi-human (QPQH) Algorithm}
We add a general pseudo-code integrating all algorithms for QPQH. For the initialization, we set $R$ the current known minimum radius. Then we do the following iterations until reaching a time limit $t_1$ or $R$ is not improved at the last iteration. At each iteration, we first call the Global Search Algorithm GSA($R$); if GSA returns a feasible layout $X$, then we call the Container Adjustment Algorithm CAA$(X, R)$ and wish it output a smaller $R$ with a new feasible layout $X$. The pseudo-code of QPQH is shown in Algorithm \ref{alg:QPQH}.

\begin{algorithm}[!htbp]
\small
\caption{The efficient Quasi-physical Quasi-human algorithm}
\label{alg:QPQH}
\begin{algorithmic}[1]
\REQUIRE ~~\\
    the current known minimum container radius $R_{bestknown}$
\ENSURE ~~\\
    The feasible minimum layout $(X,R)$ or NULL.
\STATE $R \leftarrow R_{bestknown}$;
\STATE $R_{best} \leftarrow 0$;
\WHILE {the time limit $t_1$ is not reached }
    \STATE $X \leftarrow GSA(R)$;
    \IF {$X$ is feasible}
        \STATE $R' \leftarrow R$;
        \STATE $(X, R) \leftarrow CAA(X, R)$;
            \IF {$R = R'$}
               \RETURN the current layout $(X,R)$;
            \ENDIF
            \IF {$R < R'$}
                \STATE $X_{best} \leftarrow X$;
                \STATE $R_{best} \leftarrow R$;           
            \ENDIF
    \ENDIF
\ENDWHILE
\IF {$R_{best} > 0$}
   \RETURN the best layout $(X_{best},R_{best})$;
\ENDIF
\RETURN NULL.
\end{algorithmic}
\end{algorithm}

\section{Experimental Results}
\label{Res}
In this section, we introduce the experimental environment, the key parameter tuning,  the experimental instances and results. The experimental results are divided into two parts. One is for the first 100 cases, whether QPQH can always stably find the current best-known layouts. The other is for $n \leq 320$ we compare the best finding of QPQH to the current best-known layouts.

\subsection {Experimental setup and parameter tuning}
We implement QPQH in the C++ programming language, and the IDE is Visual C++ 6.0. We run QPQH on the Ali cloud platform (http://www.aliyun.com). The platform has an 8 core processor and an 8GB memory. The OS is Windows Server 2012, and the experiment is completed without any parallelism technique.

\begin{figure*}[htbp]
    \subfigure{
    \begin{minipage}[t]{0.45\linewidth}
      \centering
      \includegraphics[width=1\textwidth]{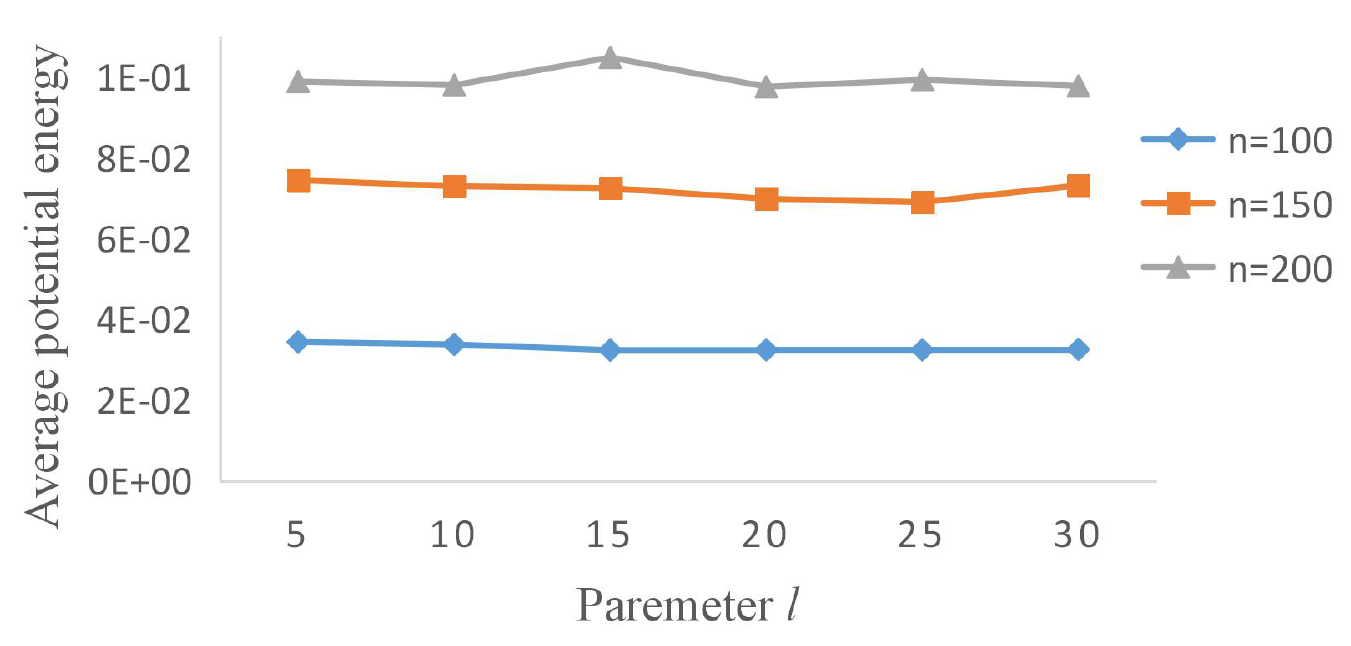}
    \end{minipage}}
    \subfigure{
    \begin{minipage}[t]{0.45\linewidth}
      \centering
      \includegraphics[width=1\textwidth]{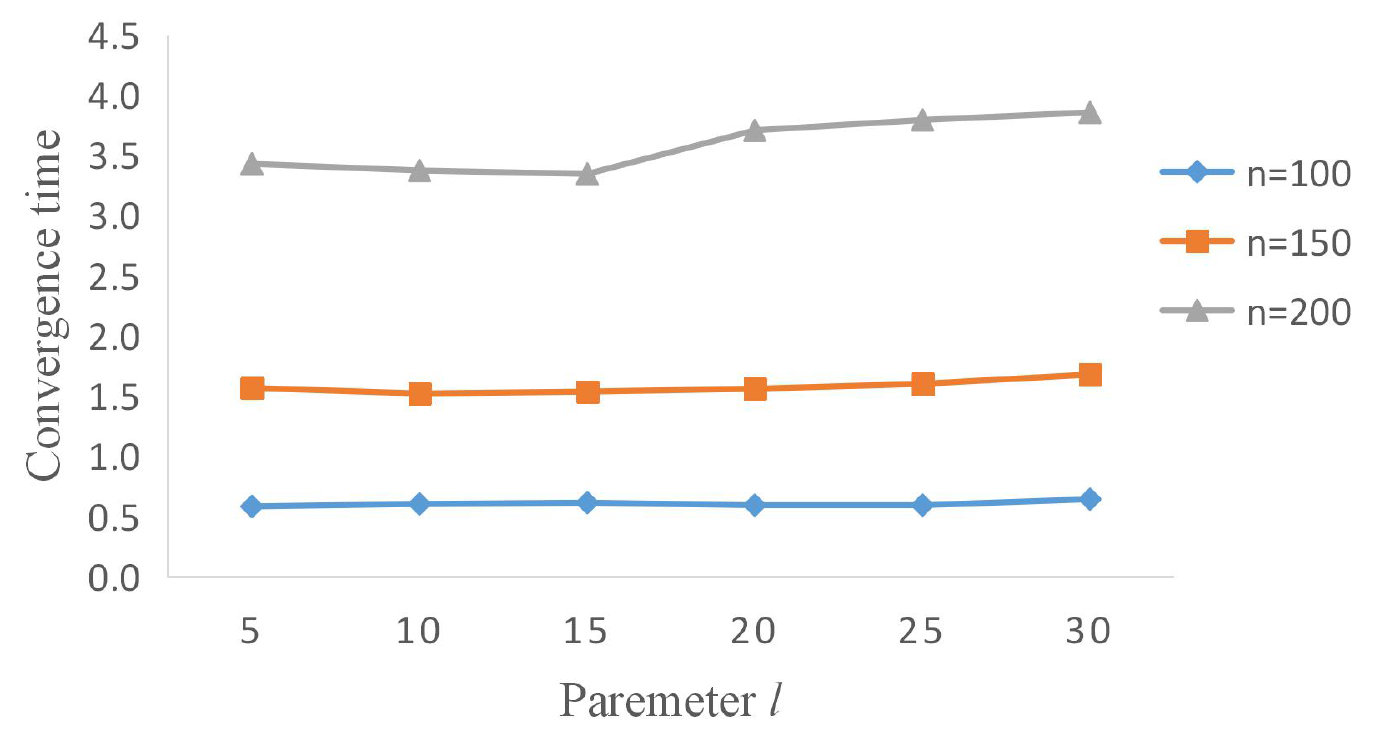}
    \end{minipage}}
\caption{Comparation of the average performance and the convergence time on different values of parameter $l$ for $n=100, 150, 200$.}
\label{fig.15}
\end{figure*}

For the key parameters in the algorithm, we pick several candidate sets of values, and did a small trial of experiments to determine the suitable setting. For example for $l$ in the local BFGS algorithm, we set $n$ = 100, 150 or 200, fix the radius to the current-best-known, and run the local BFGS algorithm for 1000 times to reach a local minima. The average potential energy and the convergence time are as shown in Figure \ref{fig.15}. We could see that the proposed algorithm is not very sensitive to the parameter $l$ as the fluctuation is very small.  We choose $l=10$ as it has a slightly shorter convergence time and slightly better average potential energy.

%

\subsection {Computational efficiency and robustness}
The first set of experiments is to test whether the algorithm can robustly find the current best-known layout. For the first 100 instances $(n = 1, 2, \cdots , 100)$, the radius of the container is set to the current known minimum (downloaded from the official website http://www.packomania.com, 20 June 2016). We randomly locate $n$ circles in the container and then call the local BFGS algorithm. Then we get 20 new layouts for the next iteration. Once the program finds a feasible layout or the running time is more than 4 hours, the calculation will terminate. In order to verify the robustness of the proposed algorithm, each instance is calculated for 10 times.

The results show that for the 100 instances $(n=1,2,\cdots, 100)$, except $n=82,100$, QPQH finds 98 current best-known layouts. For $n \leq 49$, the algorithm can always stably find the current best-known layout. In these 49 instances, the average computation time is at most 720s.

The corresponding statistics for $n=50,51,\cdots,100$ are as shown in Table \ref{table:1}. The first column is the number of  circles. The second column corresponds to the current known minimum radius of the container. The third column is the hit count among the 10 times of calculation, i.e., the number of feasible layouts found by the Global Search Algorithm (GSA) within 4 hours. And the forth column is the average computation time in seconds.
\begin{table}[H]
\setlength{\abovecaptionskip}{10pt}
\setlength{\belowcaptionskip}{0pt}
\scriptsize
\centering
\begin{tabular}{cccr|cccr}
\hline
$n$ &$R_{bestknown}$ &Hit Count &Time (s) &$n$ &$R_{bestknown}$ &Hit Count &Time (s)\\
\hline
50	&7.9475152747	&10/10	&34   &	76	&9.7295968021	&10/10	&464\\
51	&8.0275069524	&10/10	&21  &	77	&9.7989119245	&10/10	&147\\
52	&8.0847171906	&10/10	&6    &	78	&9.8577098998	&3/10	&4276\\
53	&8.1795828268	&10/10	&1043&	79	&9.9050634676	&10/10	&5464\\
54	&8.2039823834	&10/10	&33&	80	&9.9681518131	&5/10	&5145\\
55	&8.2111025509	&10/10	&27&	81	&10.0108642412	&7/10	&5446\\
56	&8.3835299225	&10/10	&21&	82	&10.0508242234	&0/10	&- \\
57	&8.4471846534	&10/10	&7&	83	&10.1168578751	&7/10	&4821\\
58	&8.5245537701	&10/10	&71&	84	&10.1495308672	&8/10	&5364\\
59	&8.5924999593	&10/10	&67&	85	&10.1631114658	&10/10	&377\\
60	&8.6462198454	&10/10	&13&	86	&10.2987010531	&2/10	&5867\\
61	&8.6612975755	&10/10	&4&	87	&10.3632085050	&10/10	&1148\\
62	&8.8297654089	&10/10	&861&	88	&10.4323376927	&10/10	&4866\\
63	&8.8923515375	&10/10	&92&	89	&10.5004918145	&5/10	&7751\\
64	&8.9619711084	&10/10	&2872&	90	&10.5460691779	&10/10	&588\\
65	&9.0173973232	&10/10	&34&	91	&10.5667722335	&10/10	&40\\
66	&9.0962794269	&10/10	&2404&	92	&10.6846458479	&4/10	&4992\\
67	&9.1689718817	&10/10	&7492&	93	&10.7333526002	&3/10	&6092\\
68	&9.2297737467	&10/10	&3252&	94	&10.7780321602	&8/10	&5552\\
69	&9.2697612666	&10/10	&713&	95	&10.8402050215	&5/10	&6056\\
70	&9.3456531940	&10/10	&1752&	96	&10.8832027597	&4/10	&5824\\
71	&9.4157968968	&10/10	&916&	97	&10.9385901100	&1/10	&11540\\
72	&9.4738908567	&10/10	&113&	98	&10.9793831282	&9/10	&2593\\
73	&9.5403461521	&2/10	&5571&	99	&11.0331411514	&9/10	&3445\\
74	&9.5892327643	&6/10	&5343&	100	&11.0821497243	&0/10	&-\\
75	&9.6720296319	&8/10	&6671\\				
\hline
\end{tabular}
\caption{The bestknown container radius, hit count and average running time for 10 rounds on $n=50,51,\cdots,100$.}
\label{table:1}
\end{table}
%
%
The second set of experiments is to test whether QPQH can find a better layout than the current best. For the first 320 instances $(n=1, 2, \cdots, 320)$, the radius of the container is set to the current known minimum (downloaded from official website http://www.packomania.com, 20 June 2016). We randomly locate $n$ circles in the container and run GSA (the time limit $t_0$ is 4 hours). Once we find the minimum layout, we apply the container adjustment procedure to decrease the radius of the container, and run QPQH again, until we can not reduce the container size within the time limit $t_1$ which is generally set to 3 days.

The results show that for 66 of the 320 instances, QPQH finds better layouts than the current best. Table \ref{table:3} shows the specific improvements.
In Table \ref{table:3}, $R_0$ is the current known minimum container radius, $R^*$ is the container radius obtained by QPQH, and $R_0-R^*$ is the improvement degree. The improvements are in the range of $10^{-7}$ to $10^{-2}$, and most improvements are $10^{-3}$ or $10^{-4}$.

Figure 9 further illustrates some new and better layouts obtained by QPQH.

\begin{figure*}[htbp]
    \subfigure[$n$ = 126]{
    \begin{minipage}[t]{0.49\linewidth}
      \centering
      \includegraphics[width=0.8\textwidth]{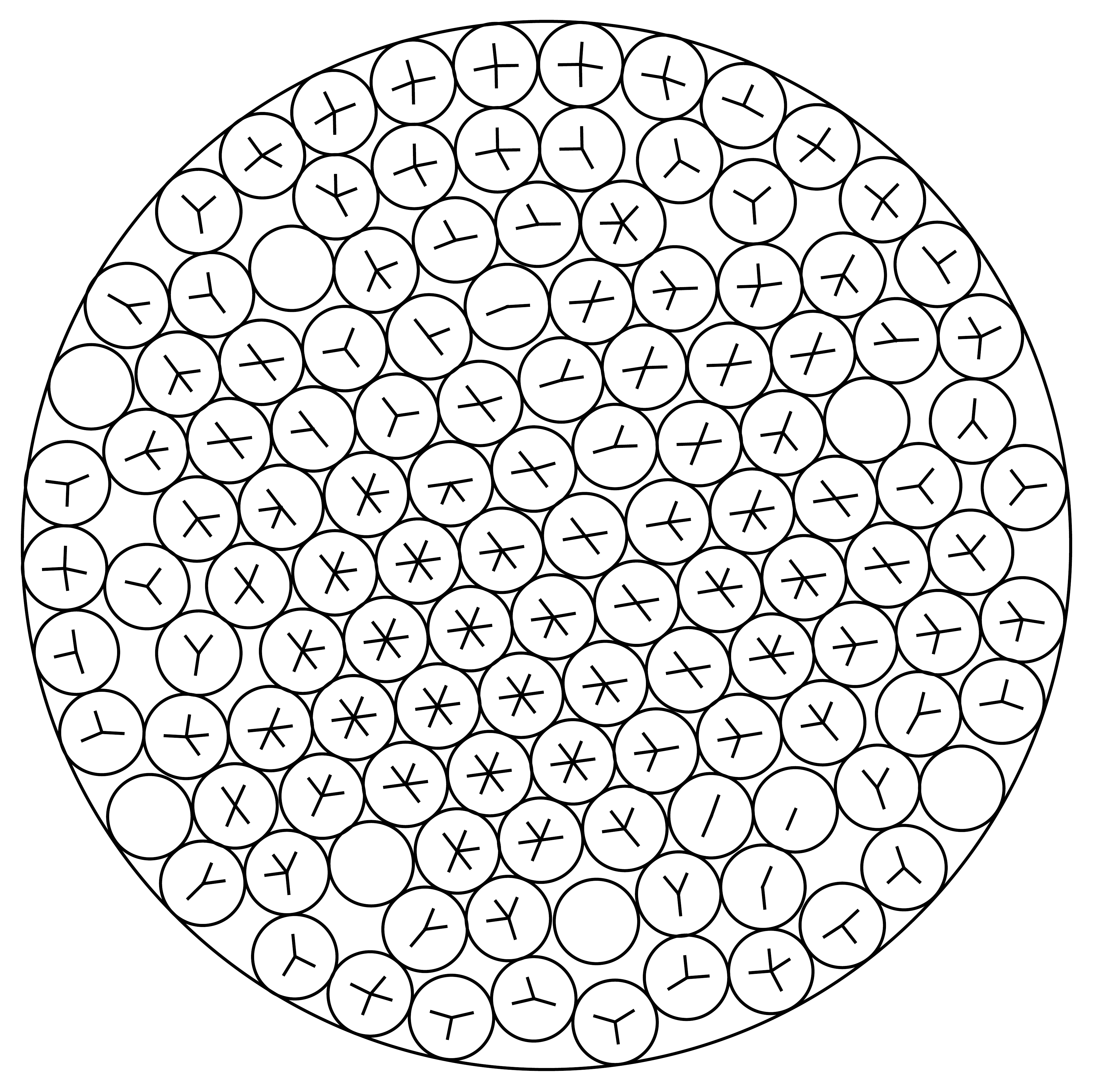}
    \end{minipage}}
    \subfigure[$n$ = 128]{
    \begin{minipage}[t]{0.49\linewidth}
      \centering
      \includegraphics[width=0.8\textwidth]{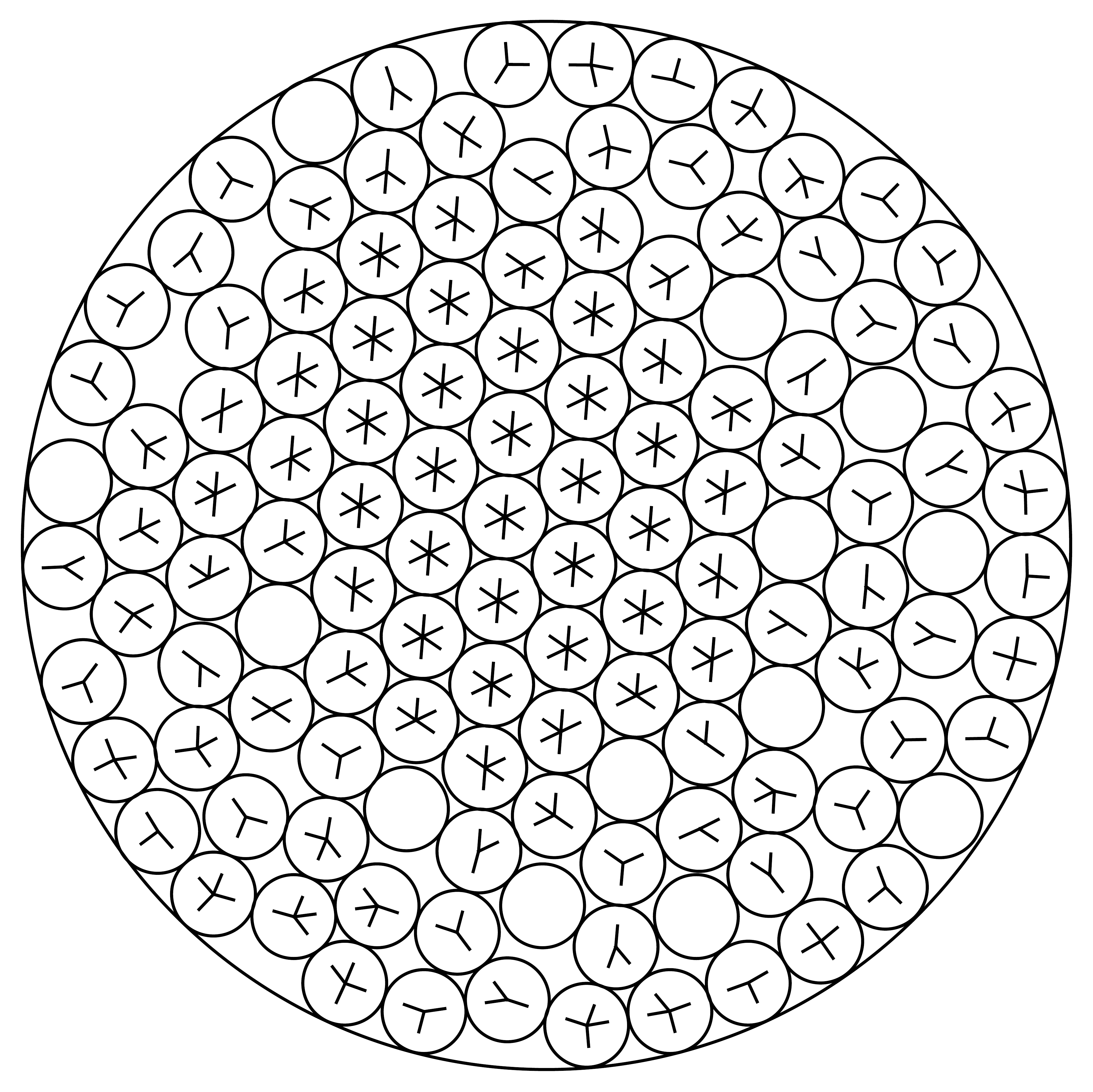}
    \end{minipage}}
    
    \subfigure[$n$ = 130]{
    \begin{minipage}[t]{0.49\linewidth}
      \centering
      \includegraphics[width=0.8\textwidth]{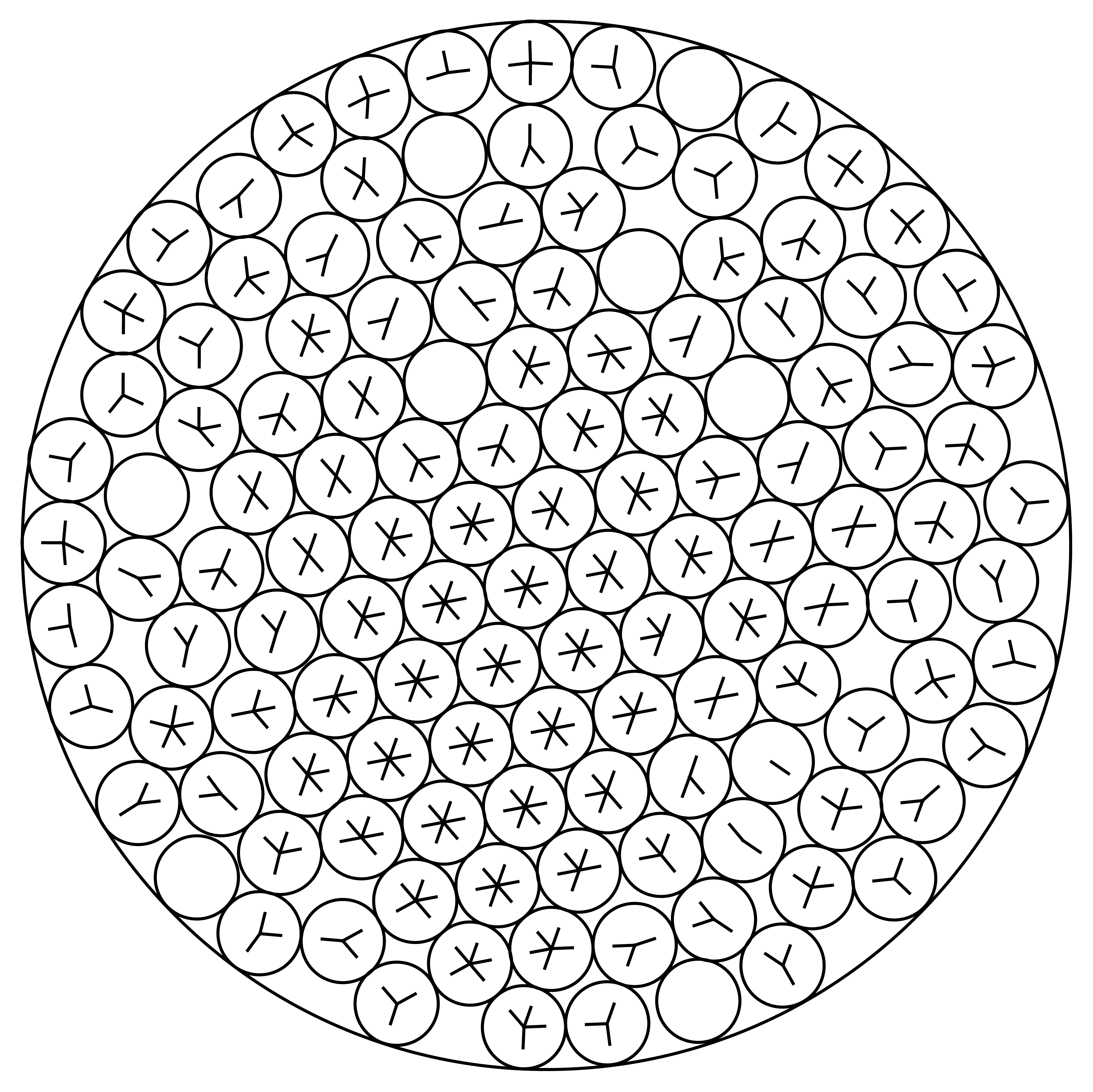}
    \end{minipage}}
    \subfigure[$n$ = 137]{
    \begin{minipage}[t]{0.49\linewidth}
      \centering
      \includegraphics[width=0.8\textwidth]{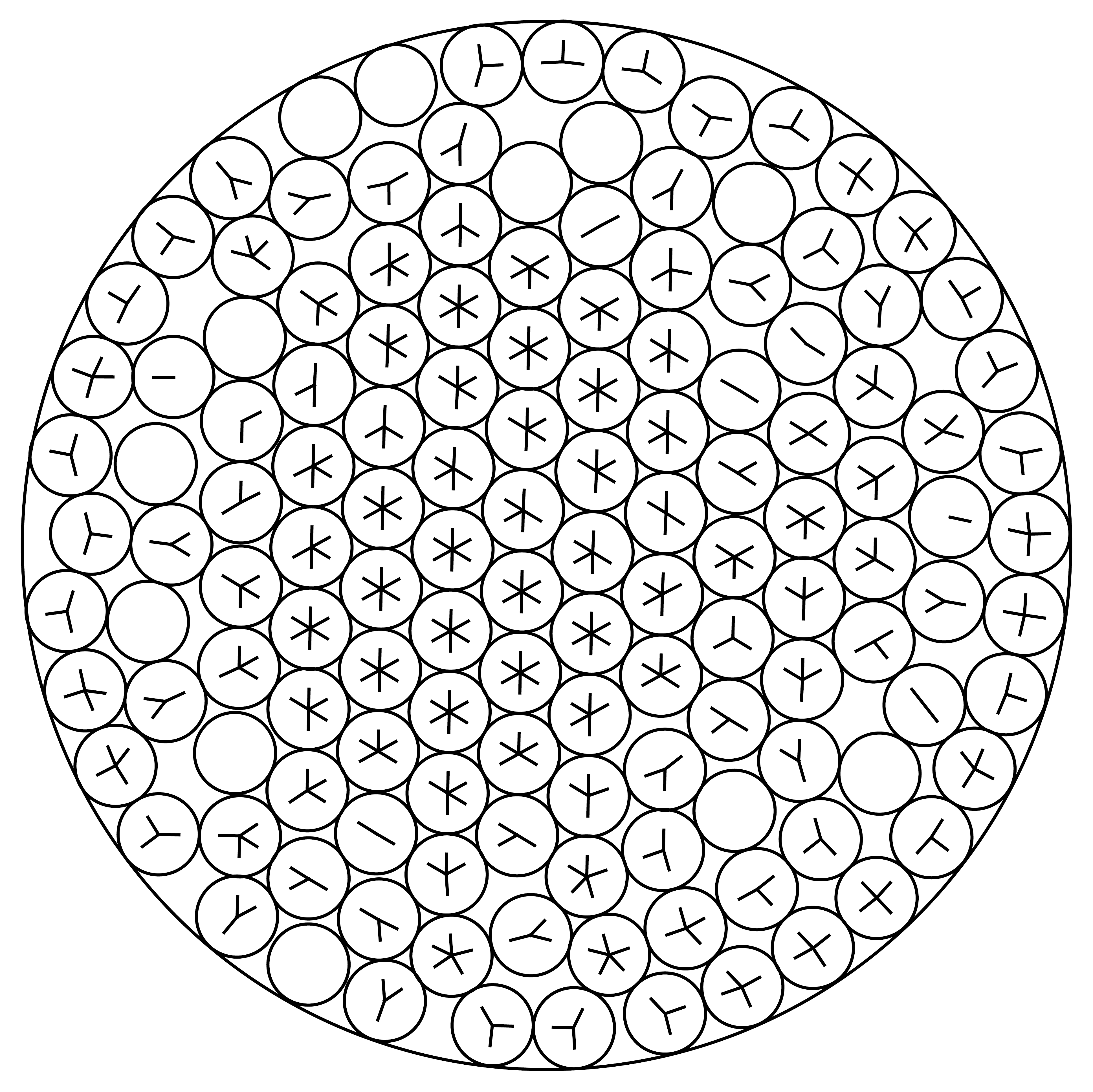}
      
    \end{minipage}}
    \subfigure[$n$ = 138]{
    \begin{minipage}[t]{0.49\linewidth}
      \centering
      \includegraphics[width=0.8\textwidth]{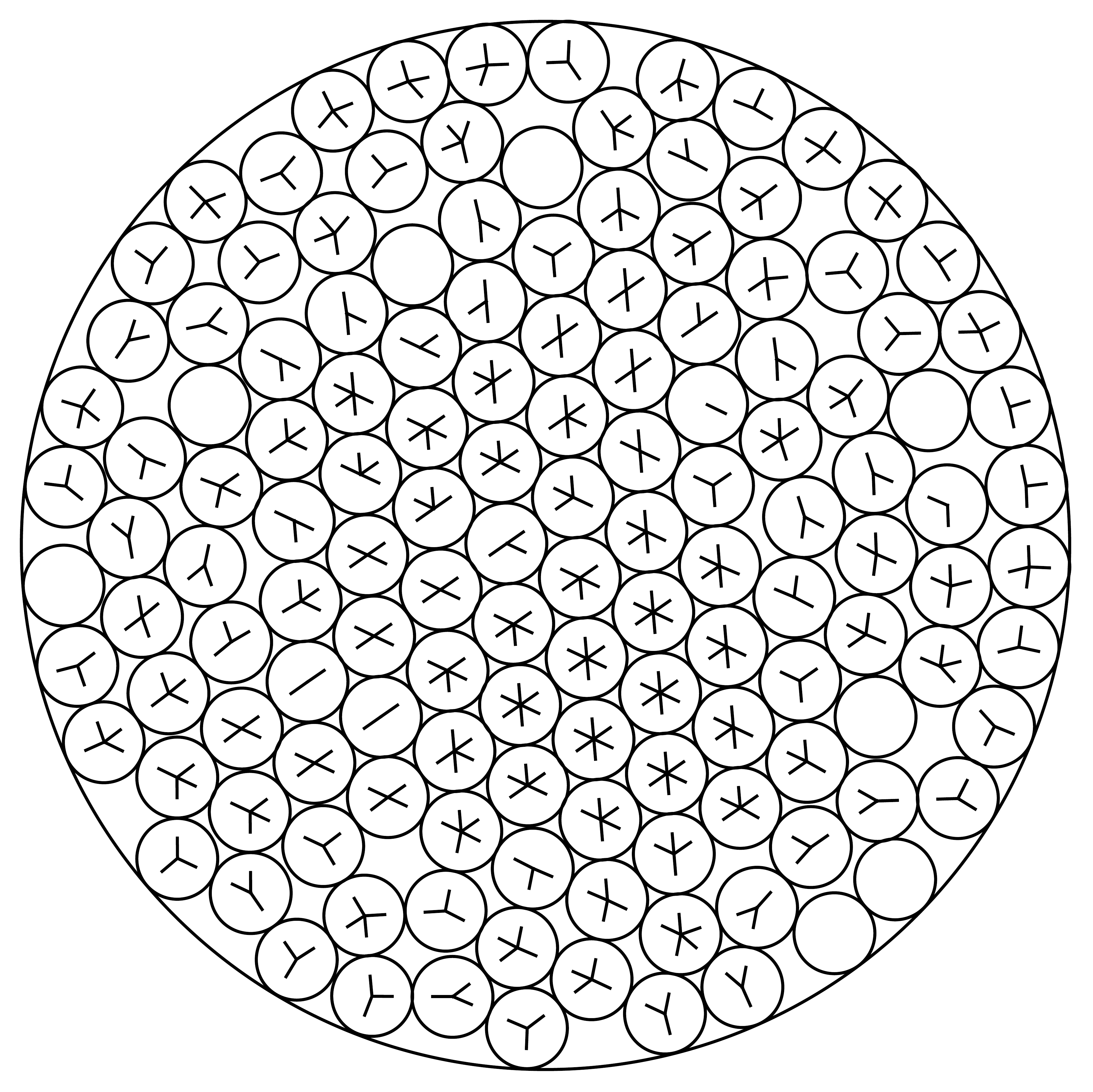}
    \end{minipage}}
    \subfigure[$n$ = 140]{
    \begin{minipage}[t]{0.49\linewidth}
      \centering
      \includegraphics[width=0.8\textwidth]{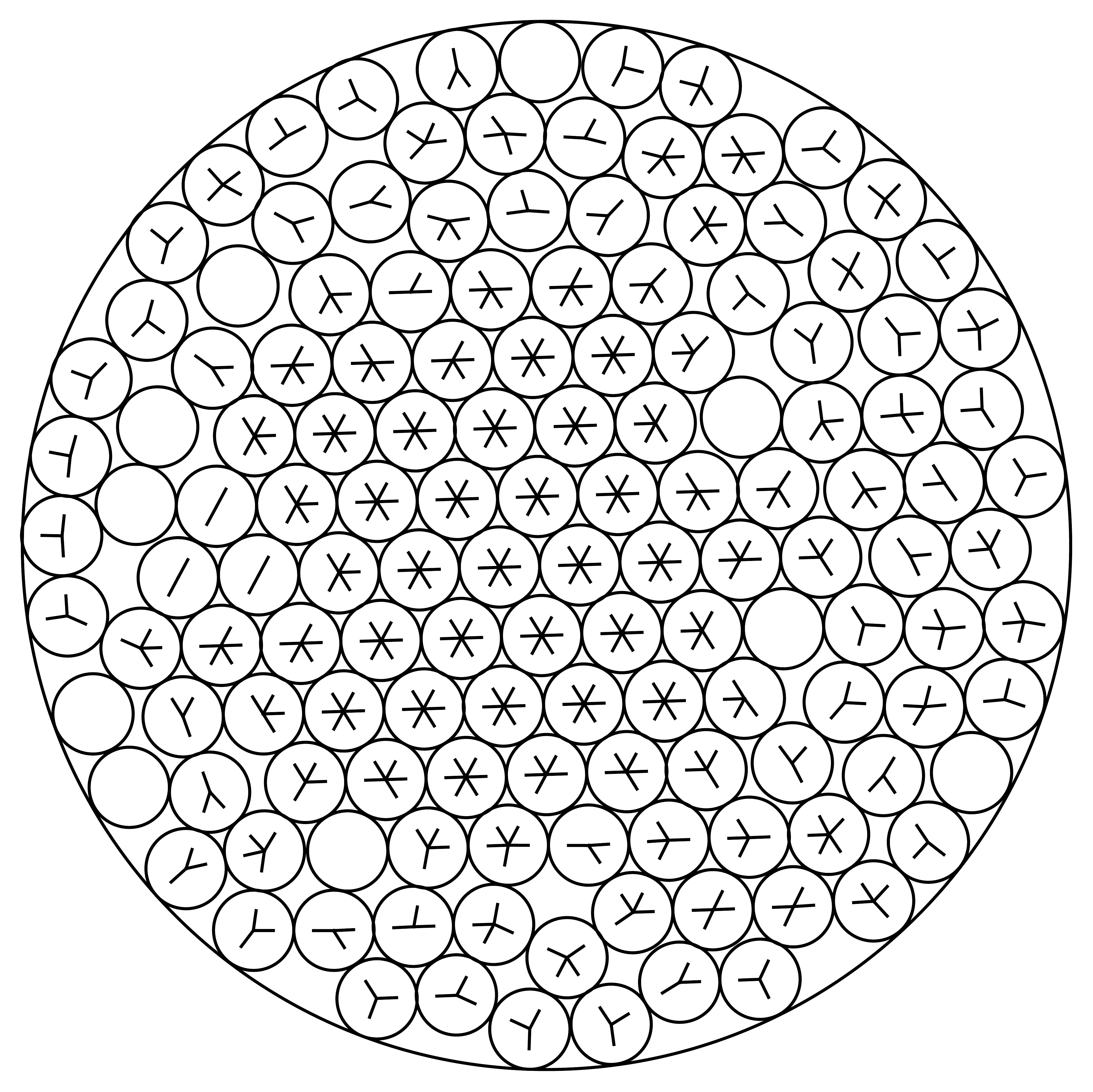}
    \end{minipage}}
\caption{New and better layouts abtained by QPQH on some instances.}
\label{fig.6}
\end{figure*}

\begin{table}[H]
\setlength{\abovecaptionskip}{10pt}
\setlength{\belowcaptionskip}{0pt}
\scriptsize
\centering
\begin{tabular}{cccc|cccc}
\hline
$n$ &$R_0$ &$R^*$ &$R_0-R^*$ &$n$ &$R_0$ &$R^*$ &$R_0-R^*$\\
\hline
126	&12.417463955   &12.417144417	&$10^{-4}$	&230	&16.596430072	&16.596246697	&$10^{-4}$	\\
128	&12.50231007	&12.502222199	&$10^{-4}$	&231	&16.640078326	&16.631031907	&$10^{-2}$	\\
130	&12.602318936	&12.601774612	&$10^{-3}$	&242	&16.962132986	&16.961287246	&$10^{-3}$	\\
137	&12.914725416	&12.914711247	&$10^{-5}$	&243	&17.004065249	&17.001947599	&$10^{-3}$	\\
138	&12.962702608	&12.961304417	&$10^{-3}$	&244	&17.039719463	&17.039559451	&$10^{-4}$	\\
140	&13.061097215	&13.060696617	&$10^{-4}$	&245	&17.079365817	&17.078003394	&$10^{-3}$	\\
156	&13.719600039	&13.718404613	&$10^{-3}$	&246	&17.113998222	&17.113112319	&$10^{-3}$	\\
160	&13.920538614	&13.920451761	&$10^{-4}$	&247	&17.141082423	&17.132526683	&$10^{-2}$	\\
163	&14.069620216	&14.067635971	&$10^{-3}$	&248	&17.184447103	&17.182444113	&$10^{-3}$	\\
177	&14.617194154	&14.614803186	&$10^{-3}$	&251	&17.305245615	&17.297607156	&$10^{-2}$	\\
178	&14.658814223	&14.655927628	&$10^{-3}$	&252	&17.331245569	&17.326883903	&$10^{-2}$	\\
179	&14.702293364	&14.699982808	&$10^{-3}$	&254	&17.400734608	&17.400641308	&$10^{-4}$	\\
180	&14.742878035	&14.739035484	&$10^{-3}$	&255	&17.454200693	&17.454058463	&$10^{-4}$	\\
183	&14.869399059	&14.865788757	&$10^{-3}$	&256	&17.494310641	&17.49310149	&$10^{-3}$	\\
188	&15.028782734	&15.028750763	&$10^{-5}$	&258	&17.547880529	&17.547085193	&$10^{-3}$	\\
194	&15.249753585	&15.241428667	&$10^{-2}$	&261	&17.634862232	&17.634766878	&$10^{-4}$	\\
197	&15.368975294	&15.36851843	&$10^{-4}$	&277	&18.142486517	&18.139629995	&$10^{-3}$	\\
198	&15.391207855	&15.390410683	&$10^{-3}$	&278	&18.189804499	&18.187675017	&$10^{-3}$	\\
204	&15.652649998	&15.644738394	&$10^{-2}$	&279	&18.224068629	&18.221572693	&$10^{-3}$	\\
205	&15.709043665	&15.695538521	&$10^{-2}$	&281	&18.284504843	&18.281539686	&$10^{-3}$	\\
207	&15.770271663	&15.770190573	&$10^{-4}$	&282	&18.315754345	&18.309334921	&$10^{-2}$	\\
213	&15.970256294	&15.969855988	&$10^{-4}$	&286	&18.434315259	&18.434157792	&$10^{-4}$	\\
214	&16.01876322	&16.018386421	&$10^{-4}$	&296	&18.704963368	&18.703338597	&$10^{-3}$	\\
216	&16.087407652	&16.087017095	&$10^{-4}$	&301	&18.843979273	&18.843675327	&$10^{-4}$	\\
217	&16.119370348	&16.118237367	&$10^{-3}$	&302	&18.895682961	&18.89237664	&$10^{-3}$	\\
221	&16.261873984	&16.261873763	&$10^{-7}$	&303	&18.936286113	&18.935589502	&$10^{-3}$	\\
222	&16.299696226	&16.299162417	&$10^{-3}$	&304	&18.973327182	&18.972060389	&$10^{-3}$	\\
224	&16.37189924	&16.369591221	&$10^{-3}$	&305	&19.010908936	&19.008914111	&$10^{-3}$	\\
225	&16.409054279	&16.408410528	&$10^{-3}$	&308	&19.121680199	&19.120984457	&$10^{-3}$	\\
226	&16.450019501	&16.449824611	&$10^{-4}$	&318	&19.407004242	&19.396538989	&$10^{-2}$	\\
227	&16.494400009	&16.489753739	&$10^{-2}$	&319	&19.432992566	&19.432246442	&$10^{-3}$	\\
228	&16.52734087	&16.527071189	&$10^{-4}$	&320	&19.456230764	&19.451309906	&$10^{-2}$	\\
229	&16.566377981	&16.564350195	&$10^{-3}$	\\		
\hline
\end{tabular}
\caption{Improvements in the case of $n=126,127,\cdots,320$.}
\label{table:3}
\end{table}

\section{Conclusion}
\label{Con}
We propose an efficient Quasi-physical Quasi-human (QPQH) algorithm for solving the classical Equal Circle Packing Problem (ECPP). Our contributions include:	
(1) We modified the classic Quasi-Newton method BFGS by only considering the neighbor structure of each circle during the iteration process. In this way, we considerably speed up the calculation, especially for large scale instances.
(2) We propose a new basin-hopping strategy of shrinking the container size and the container adjustment procedure, which is simple but very effective. 
Experiments on 320 ECPP instances ($n=1, 2, \cdots, 320$) demonstrate the success of the proposed algorithm. For the first 100 instances ($n=1, 2, \cdots, 100$), we found 98 current best-known layouts  except $n=82,100$, and we found 66 better layouts for all the 320 instances. In the future, we will adapt the idea of solving ECPP for unequal circles packing problem.



	

%
%
%

\bibliographystyle{elsarticle-harv}
\bibliography{reference_packing}
\end{document}